\documentstyle[prd,aps,epsfig]{revtex}

\begin{document}

\title{Conservation Laws
and Particle Production in Heavy Ion Collisions\footnote{Invited
talk presented at International Symposium on Hadrons and Nuclei,
Seoul, Korea, 2001. }}

\author{K. Redlich$^{1,2}$, J. Cleymans$^3$, H. Oeschler$^4$, A.
Tounsi$^5$}

\address{$^1$\ Gesellschaft f\"ur Schwerionenforschung, D-64291 Darmstadt,
Germany}

\address{$^2$\ Institute for Theoretical Physics, University of Wroclaw,
PL-50204  Wroclaw, Poland  }

\address{$^3$\ Department of Physics, University of Cape Town,
Rondebosch 7701, Cape Town, South Africa}

\address{$^4$\ Institut f\"ur Kernphysik, Darmstadt University of
Technology, D-64289~Darmstadt, Germany}

\address{$^5$\ Laboratoire de Physique Th\'eorique et Hautes Energies,
 Universit\'e Paris 7,  F--75251 Cedex 05, France }

\maketitle

\noindent
\begin{abstract}
We discuss the role of the conservation laws related with U(1)
internal symmetry group in the statistical model description of
particle productions in ultrarelativistic heavy ion collisions. We
derive and show the differences in particle multiplicities in the
canonical and the grand canonical formulation of quantum number
conservation. The time evolution and the approach to chemical
equilibrium in the above ensembles   is discussed in terms of
kinetic master equation. The application of the statistical model
to the description of (multi)strange particle yields  at GSI/SIS
and the SPS energies is also presented.
\end{abstract}


\bigskip

\section{Introduction}

There are in general two approaches to describe integrated
particle yields measured in ultrarelativistic heavy ion
collisions: (i) the microscopic transport models \cite{trans} and
(ii) the macroscopic statistical thermal models
\cite{stat,Bra99,CLK,Bra95,becattini2}. In this article we will
discuss the statistical approach and show that it provides a very
satisfactory description of experimental data. We will emphasize
the importance of the conservation laws in the particular
strangeness conservation when modelling particle chemical
freezeout conditions.

Within a statistical approach, the production of particles is
commonly described  using the grand canonical (GC) ensemble, where
the charge conservation is controlled by the related chemical
potential. In this description a net value of a given U(1) charge
is conserved on the average. The (GC) approach can be only valid
if the total number of particles carrying quantum number related
with this symmetry is very large. In the opposite limit of a small
particle multiplicities, conservation laws must be implemented
exactly and locally, i.e., the canonical (C) ensemble for
conservation laws must be used \cite{Hag71,ko}. The local
conservation of quantum numbers in the canonical approach severely
reduces the phase space available for particle productions. This
treatment of charge conservation is of crucial importance in the
description of particle multiplicities in proton induced processes
\cite{b1,hamieh}, in $e^+e^-$ \cite{b1} as well as in central
heavy ion collisions at low beam energies \cite{Cle99}.

In this article      we   describe   the  exact strangeness
conservation in the context of relativistic statistical
thermodynamics. A kinetic theory for the time evolution of
particle production and the approach to the grand canonical  and
the canonical  equilibrium distribution will be also introduced.
Finally the example of the applications of the statistical model
in (C) ensemble is presented in the context of low energy central
as well as in high energy peripheral    heavy ion collisions.

\section{Statistical model and particle multiplicity}
The exact treatment of quantum numbers in statistical mechanics
has been well established for some time now \cite{Hag71,ko,Cle99}.
It is in general obtained by projecting the partition function
onto the desired values of the conserved charges  by using  group
theoretical methods.  For our purpose we shall only consider the
conservation laws related to the abelian U(1) symmetry group. In
particular, we concentrate on strangeness conservation.

The basic quantity in the statistical mechanics describing a
thermal properties of a system is the partition function $Z(T,V)$.
In the (GC) ensemble,

\begin{equation}
Z^{GC}(T,V,\mu_Q)\equiv Tr[e^{-\beta (H -\mu_Q Q)}]
\end{equation}
where $Q$ is the  conserved charge, $H$ the hamiltonian of the
system, $\mu_Q$  is the chemical potential which plays  the role
of the Lagrange multiplier   which guarantees that the charge $Q$
is conserved on the average in the whole system. Finally $\beta
=1/T$ is the inverse temperature.

In the (C) ensemble the charge $Q$ is conserved exactly. Thus
there is no  more   chemical potential under the trace and instead
we  calculate  the partition function summing only these states
which are carrying  exactly the quantum number $Q$, that is

\begin{equation}
Z_Q^{C}(T,V)\equiv Tr_Q[e^{-\beta H }]
\end{equation}

The canonical and the grand canonical partition functions are
related through the following cluster decomposition,

\begin{equation}
Z^{GC}(T, V, \lambda )= \sum_{Q=-\infty }^{+\infty} \lambda^Q
Z_Q^{C}(T,V)
\end{equation}
where the fugacity $\lambda\equiv \exp(\beta\mu_Q)$  and  the sum
is taken over all possible values of the charge $Q$.

%
 For the   (GC) partition function, which is well behaving analytic
function of the fugacity $\lambda$, the above relation can be
inverted and the canonical partition function with a given value
of the charge  $Q$ reads,

\begin{equation}
Z_Q={1\over {2\pi }}\int_0^{2\pi}d\phi e^{-iQ\phi } {\tilde
Z}(T,V,\phi ) \label{e:zq}
\end{equation}
\noindent where the generating function $\tilde Z$ is obtained
from the grand canonical partition function replacing the fugacity
parameter $\lambda$ by the factor $e^{i\phi }$,
\begin{equation}
{\tilde Z}(T,V,\phi )\equiv
       Z^{GC}(T,V,\lambda \to e^{i\phi })
\end{equation}
The  form of the generating function $\tilde Z$ in the above
equation is model dependent. Having in mind the application of the
statistical description to particle production  in heavy ion
collisions   we calculate $\tilde Z$ in the ideal gas
approximation, however, including all particles and resonances
\cite{Cle99}. This is not an essential restriction, because,
describing the freeze-out conditions we are dealing with a dilute
system where the interactions should not influence particle
production anymore. We neglect any medium effects on particle
properties. In general, however, already in the low-density limit,
the modifications of resonance width  or particle dispersion
relations  could be of importance \cite{medium}. For the sake of
simplicity, we use classical    statistics, i.e. we assume
temperature and density regime so that all particles can be
treated  using Boltzmann distributions.

In  nucleus-nucleus collisions  the absolute values of the baryon
number, electric charge and strangeness are fixed by the initial
conditions. Modelling particle production in statistical
thermodynamics would in general require the canonical formulation
of all these quantum numbers. From the previous analysis
\cite{Hag71,Cle99}, however, it is clear, that in heavy ion
collisions only strangeness should be treated exactly, whereas the
conservation of baryon and electric charges can be described by
the appropriate chemical potentials in the grand canonical
ensemble.

Within the approximations described above and neglecting first the
contributions from multi-strange baryons, the generating function
in  equation  (\ref{e:zq})  has  the  following  form,
\begin{equation}
{\tilde Z}(T,V,\mu_Q,\mu_B,\phi )=\exp (N_{s=0}+N_{s=1}e^{i\phi
}+N_{s=-1} e^{-i\phi }) \label{e:zs0}
\end{equation}
\noindent where $N_{s=0,\pm 1}$ is defined as the sum over all
particles and resonances
 having
strangeness $0,\pm 1$,
\begin{equation}
N_{s=0,\pm 1}=\sum_k Z_k^1 \label{e:N}
\end{equation}
\noindent and  $Z^1_k$ is the one-particle partition function
defined as
\begin{equation}
Z_k^1\equiv {{Vg_k}\over {2\pi^2}} m_k^2TK_2(m_k/T)\exp
(b_k\mu_B+q_k\mu_q) \label{e:zk1}
\end{equation}
\noindent with the mass $m_k$, spin-isospin degeneracy factor
$g_k$,   particle baryon number $b_k$ and electric charge $q_k$.
The volume of the system is $V$ and the chemical potentials
related with the electric charge and the baryon number  are
determined by $\mu_q$ and $\mu_B$ respectively.

With the particular form  of the generating function equations
(\ref{e:zs0},\ref{e:N},\ref{e:zk1}) the $\phi$-integration in
equation (\ref{e:zq})  can be done analytically giving the
canonical partition function for a gas with total strangeness $S$
\cite{Cle99}:
\begin{equation}
Z_{S}(T,V,\mu_B,\mu_Q)= Z_{0}(T,V,\mu_B,\mu_Q) I_S(x)
\end{equation}
\noindent where $Z_0=\exp {(N_{S=0})}$ is the partition function
of all particles having zero strangeness and the argument of the
Bessel function
\begin{equation}
x\equiv 2\sqrt {S_1S_{-1}}. \label{e:x}
\end{equation}
\noindent with $S_{\pm 1}\equiv N_{s=\pm 1}$. The parameter $x$
thus measures the total number of strange particles in thermal
fireball.

The calculation of the particle density $n_k$ in the canonical
formulation is straightforward. It amounts to the replacement
\begin{equation}
Z_k^1 \mapsto \lambda_k Z_k^1
\end{equation}
of the corresponding one-particle partition function  in equation
(\ref{e:N}) and taking the derivative of the canonical partition
function equation (\ref{e:zq}) with respect to $\lambda_k$

\begin{equation}
n_k^C\equiv [\lambda_k  {{\partial}\over
 {\partial\lambda_k}}\ln Z_Q(\lambda_k) ]_{\lambda_k=1}
\end{equation}

As an example, we quote the result for the density of thermal
kaons in the canonical formulation assuming  that the total
strangeness of the system $S=0$,
\begin{equation}
n_{K}^C={{Z_{K}^1}\over V} {{S_1}\over {\sqrt {S_1S_{-1}}}}
{{I_{1}(x)}\over {I_0(x)}} \label{e:nkc1}
\end{equation}
Comparing the above formula with the result for thermal kaons
density in the grand canonical  ensemble, $n_K^{GC}=(Z_K^1/V)\exp
{(\mu_s/T)}$, one can see that the canonical result can be
obtained from the grand canonical one replacing the  strangeness
fugacity $\lambda_S \equiv \exp {(\mu_S/T)}$ in the following way:
\begin{equation}
n_K^{C}=n_K^{GC}\left(\lambda_S\mapsto {{S_1}\over {\sqrt
{S_1S_{-1}}}} {{I_1(x)}\over {I_0(x)}}\right) \label{e:nkc2}
\end{equation}
In the limit of large $x$ that is large volume and/or temperature
 the canonical and the grand canonical formulation are
equivalent. For a small number of strange particles in a system,
however, the differences are large. This can be seen in the most
transparent way when comparing two limiting situations: the large
and small $x$ limit of the above equation.
In the  limit $x\to \infty$ we have
\begin{equation}
\lim_{x\to \infty } {{I_1(x)}\over {I_0(x)}}\to 1 \label{e:i1i01}
\end{equation}
\noindent and the kaon     density is independent of the volume of
the system as expected in the grand canonical ensemble. On the
other hand in the limit of a small $x$ we have

\begin{equation}
\lim_{x\to 0 } {{I_1(x)}\over {I_0(x)}}\to {x\over 2}
\label{e:i1i02}
\end{equation}

\noindent and  the particle density is linearly dependent on the
volume. It is thus clear, that the major difference between the
canonical     and the grand canonical
      treatment of the conservation laws appears
through different volume dependence of strange particle densities
as well as strong suppression of thermal particle phase space.
 The relevant parameter, $F_S$, which
 measures the suppression
 of particle multiplicities from their grand canonical result
is determined by the
 ratio of the Bessel functions
\begin{equation}
F_S\equiv  {{I_1(x)}\over {I_0(x)}}
\end{equation}
\noindent with the argument $x$ defined in equation (\ref{e:x}).

In Fig.~1 we show the canonical suppression factor $F_S(x)$ as a
function of the argument $x$.
 To relate the initial volume of the system
with the number of participant in A-A collisions one uses the
approximate relation $V\sim 1.9\pi A_{part}$. The corresponding
value of $x$ at SIS, AGS and SPS energy is calculated with the
baryon chemical potential and temperature extracted from the
measured particle multiplicity ratios \cite{Cle99}. The results in
Fig.~1 shows the importance of the canonical treatment of
strangeness conservation at SIS energy. Here, the canonical
suppression factor can be even larger than
 an order of magnitude.

For central Au-Au collisions at AGS or SPS energy this suppression
is not  relevant any more and the (GC)-formalism is adequate. In
general, one expects, that the statistical interpretation of
particle production in central heavy ion collisions requires the
 canonical treatment of strangeness
conservation if the CMS collation energy $\sqrt {s} <2-3$GeV. This
is mainly because at these energies the freeze-out temperature is
still too low to maintain large argument expansion of the Bessel
functions in equation (\ref{e:i1i01}). The canonical description
of strangeness conservation can be, however, also of importance at
the SPS energy were $\sqrt {s}\sim 18$GeV  when one considers the
peripheral heavy ion collisions. This is particularly true for
multistrange particle production since  the canonical suppression
of the thermal particle phase-space increases with strangeness
content of the particle \cite{hamieh}.
\begin{figure}[htb]
\begin{minipage}[t]{170mm}
{\hskip 4.0cm
\includegraphics[width=21.5pc, height=19.5pc]{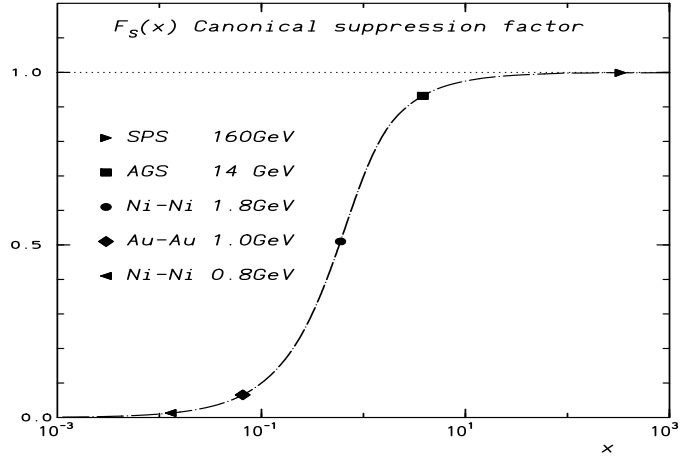}}\\
\vskip -1.0cm {\caption{Canonical strangeness suppression factor
(see text).}}
\end{minipage}
\end{figure}

\subsection{Multistrange particle multiplicities}

The extension of the canonical description to multi-strange
particle multiplicities is straightforward. One needs first to
extend the generating functional in Eq. 6 by including the
contributions of multistrange baryons. In this case the canonical
partition function constraint by the  strangeness neutrality
condition reads \cite{hamieh,esko},

\begin{equation}
Z^C_{S=0}={1\over {2\pi}}
       \int_{-\pi}^{\pi}
                   d\phi~ \exp{(\sum_{s=- 3}^3
S_se^{is\phi})}, \label{eq1}
\end{equation}
where $S_s\equiv\sum_i Z_i^1$ and the sum is taken over all
particles and resonances carrying strangeness $s$. The
one-particle partition function $Z_i^1$ is defined in Eq.(8).

With the particular
 form of the partition
function given by Eq.\,(\ref{eq1})  the  density $n_s$ of particle
$i$ with
 strangeness $s$  in volume $V$
 is obtained by the replacement  $Z_i\mapsto \lambda_iZ_i$
in Eq.\,(\ref{eq1})  and then   taking an  appropriate derivative
\cite{Hag71,hamieh}:

\begin{eqnarray}
n_{\pm s}&=&{{\langle N_{\pm s}\rangle}\over V} \equiv
[{{\lambda_i}\over {V}}{{\partial {\ln Z_{0}}}\over
 {\partial {\lambda_i}}}]_{\lambda_i=1}\simeq
Z_{\pm s} {{(S_{\pm1})^s}\over {{(S_{+1}}S_ {-1})^{s/2}}}
 \nonumber     \\
      &&\{I_s(x_1)I_0(x_2)+\sum_{m=1}^\infty
I_m(x_2)[I_{2m+|s|}(x_1)A^{m/2}+I_{2m-|s|}(x_1)A^{-m/2}]\}/Z_{S=0}
\label{eq3}
\end{eqnarray}
where
\begin{equation}
x_k\equiv 2\sqrt {S_kS_{-k}}~~,~~
 A\equiv {{S_{-1}^2S_2}\over {S_1^2S_{-2}}},
\end{equation}
and the partition function
\begin{equation}
Z_{S=0}\simeq I_0(x_1)I_0(x_2)+ \sum_{m=1}^\infty
I_{2m}(x_1)I_m(x_2)[A^{m/2}+A^{-m/2}].
\end{equation}

In the derivation of Eq.\,(\ref{eq3}) we have neglected, after
differentiation over  particle fugacity, the term $S_{\pm 3}$.
This approximation, however, due to
 small value of
$S_{\pm 3}$ coefficients in comparison with $S_{\pm 1} ~{\rm and}~
S_{\pm 2}$
 is quite satisfactory. The complete expression for the partition function
without any approximation can be found in reference \cite{esko}.

In the large system like Pb-Pb and for large collision energy,
required to reach  high $T$, the density $n_s$ of particle
carrying strangeness $s$ is $V$ independent. In the opposite
limit, however, this dependence is changed to $n_s\sim V^{s}$,
which can be verified from Eq.\,(\ref{eq3}). Indeed for small
$x_i$ we have approximately:

\begin{equation}
n_{\pm s}  \simeq Z_{\pm s} {{(S_{\pm1})^s}\over {{(S_{+1}}S_
{-1})^{s/2}}}~
      { { I_s(x_1)}\over {I_0(x_1)}}.
\label{eq6}
\end{equation}
and when expanding the Bessel functions  $I_n(x)\sim x^n$ in
Eq.\,(\ref{eq6})
 we see that
 $n_s \sim V^s$.

From the above expression  it is  clear that  strangeness
suppression, which is measured by the ration  $I_s/I_0$, is
increasing with the strangeness content of the particle. Thus,
there are two important ingredients  of canonical modifications of
multistrange particle density with respect to their (GC) value:
(i) the density is volume dependent that is also centrality
 dependent and
(ii) the thermal phase is suppressed and this suppression
increases with strangeness content of the particle.
\begin{figure}[htb]
\begin{minipage}[t]{170mm}
{\hskip 4.0cm
\includegraphics[width=21.5pc, height=19.5pc]{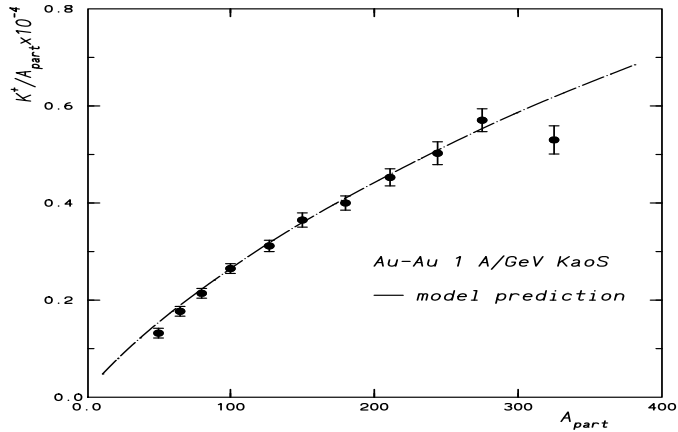}}\\
\vskip -1.0cm {\caption{Measured  $K^+$ multiplicity per
participant $A_{part}$ as a function of $A_{part}$ for Au-Au
collisions at 1 A/GeV  together with the  canonical model
results}}
\end{minipage}
\end{figure}

\section{Time evolution and strangeness equilibration}

In the last section we have formulated the statistical model for
strange particle multiplicities $<N_S>$ assuming that the system
is in thermal and chemical   equilibrium. We have shown that
dependently on the total number of strange particles there are two
distinct equilibrium limits \cite{ko,Cle99}: if $<N_s>$ is small
then we are in the canonical regime and in the opposite limit the
canonical and grand canonical description coincides. In this
section we consider the time evolution of the multiplicity of
strange particles and formulate a kinetic master equations which
distinguish  between these two equilibrium limits. For a sake of
illustration we consider a simple example of $K^+K^-$ production
in the environment of thermal pions in volume $V$ and temperature
$T$  due to the following binary process, $\pi^+\pi^-\to K^+K^-$.
We formulate for  this example  the kinetics for the time
evolution of kaon multiplicities and their approach to chemical
equilibrium .

In the standard formulation  \cite{koch}, the rate equation for
this binary process is described by the following population
equation:
\begin{eqnarray}
\frac{d<N_K>}{d\tau}={G\over V} <N_{\pi^+}> <N_{\pi^-}> - {L\over
V} <N_{K^+}>< N_{K^-}>, \label{normal1}
\end{eqnarray}
where $G \equiv \langle \sigma_G v \rangle$ and $L \equiv \langle
\sigma_L v \rangle$ give the momentum-averaged cross sections for
the gain  $\pi^+ \pi^- \rightarrow K^+ K^-$ and the loss  $K^+ K^-
\rightarrow \pi^+ \pi^-$ process respectively. The value of
$<N_K>$ represents the total number of produced kaons.

To include the possible correlations between the production of
$K^+$ and $K^-$ \cite{ko}, let us define $P_{i,j}$ as the
probability to find $i$ number of  $K^+$ {\em and} $j$ number of
$K^-$ in an event. We also denote by $P_i$ as the probability to
find $i$ number of  $K$  pairs in an event. The average number of
$K$ per event is defined as:
\begin{eqnarray}
\langle N_K \rangle =\sum_{i=0}^{\infty} iP_i. \label{2}
\end{eqnarray}

We can now write the following general rate equation for the
average kaon multiplicities:
\begin{eqnarray}
\frac{d\langle N_{K} \rangle }{d\tau}= {G\over V} \langle
N_{\pi^+} \rangle \langle N_{\pi^-} \rangle - \frac {L}{V}
\sum_{i,j} ij P_{i,j}. \label{general}
\end{eqnarray}

Due to the local conservation of quantum numbers, we have:
\begin{eqnarray}
P_{i,j}&=& P_i ~\delta_{ij}, \nonumber\\
\sum_{i,j} ij P_{i,j} &=& \sum_i i^2 P_i \equiv \langle N^2
\rangle =\langle N \rangle ^2+\langle \delta N^2 \rangle,
\label{5}
\end{eqnarray}
where $\langle \delta N^2 \rangle$ represents the event-by-event
fluctuation of the number of $K^+ K^-$ pairs. Note that we always
consider abundant $\pi^+$ and $\pi^-$  so that we can neglect the
number fluctuation of these particles and the change of their
multiplicities due to the considered processes.

Following Eqs.(\ref{general}-\ref{5}) the general rate equation
for the average number of $K^+K^-$ pairs can be written as:
\begin{eqnarray}
\frac{d\langle N_K \rangle }{d\tau}={G\over V} \langle N_{\pi^+}
\rangle \langle N_{\pi^-} \rangle - \frac {L}{V} \langle N_K^2
\rangle . \label{general2}
\end{eqnarray}
For abundant production of $K^+ K^-$ pairs where $\langle N_K
\rangle \gg 1$,
\begin{eqnarray}
\langle N_K^2 \rangle \approx \langle N_K\rangle ^2,
\end{eqnarray}
and Eq.(\ref{general2}) obviously reduces to the standard form:
\begin{eqnarray}
\frac{d\langle N_K\rangle }{d\tau}\approx \frac{G}{V} \langle
N_{\pi^+} \rangle \langle N_{\pi^-} \rangle - \frac{L}{V} \langle
N_K \rangle ^2. \label{normal2}
\end{eqnarray}
However, for rare production of $K^+ K^-$ pairs where $\langle
N_K\rangle\ll \!1$, the rate equations (\ref{normal1}) and
(\ref{normal2}) are no longer valid. We have instead
\begin{eqnarray}
\langle N_K^2 \rangle \approx \langle N_K\rangle ,
\end{eqnarray}
which reduces Eq.(\ref{general2}) to the following form \cite{ko}:
\begin{eqnarray}
\frac{d\langle N_K\rangle }{d\tau}\approx {G\over V}
 \langle N_{\pi^+} \rangle \langle N_{\pi^-} \rangle
- \frac{L}{V} \langle N_K \rangle . \label{canonical}
\end{eqnarray}
Thus, in the limit where $\langle N_K \rangle \ll 1$, the
absorption term depends on the pair number only linearly, instead
of quadratically for the limit of $\langle N_K \rangle \gg 1$.
Thus, it is clear that   the time evolutions and equilibrium
values for kaon multiplicities are obviously different in the
above limiting situations.

In the limit of large $<N_K>$ the equilibrium value for the number
of $K^+K^-$ pairs, which coincides with the multiplicity of $K^+$
and $K^-$, is obtained from Eq.(29) as,

\begin{eqnarray}
<N_K>_{\rm eq}^{\rm GC}= { {V}\over {2\pi^2}} m_K^2 T K_2 (M_K/T)
 \label{15}
\end{eqnarray}
thus, it is described by the  (GC) result with vanishing chemical
potential due to   strangeness  neutrality condition.

In the opposite limit where $\langle N_K \rangle \ll 1$, the time
evolution is described by Eq.(\ref{canonical}), which has the
following  equilibrium solution:
\begin{eqnarray}  N_{\rm eq}^{\rm C}= \left [
{ {V}\over {2\pi^2}} M_{K^+}^2T  K_2 (M_{K^+}/T) \right ] \left [
{ {V}\over {2\pi^2}} M_{K^-}^2T  K_2 (M_{K^-}/T)
 \right ].
\label{lim} \end{eqnarray}

The above equation demonstrates the locality of strangeness
conservation. With each $K^+$ the  $K^-$ is produced in the same
event in order to conserve strangeness exactly and locally. This
is the result expected from the  (C) formulation of the
conservation laws as described in the previous section.  We note
that Eq.(\ref{lim}) is just the leading term in the expansion of
the canonical result for multiplicities of particles which are
carrying $U(1)$ charges. The general expression is given by
Eq.(13) and Eq.(20).

 Comparing Eq.(32) and
Eq.(33), we first find that, for $\langle N_K \rangle \ll 1$, the
equilibrium value in the canonical formulation is far smaller than
what is expected from the grand canonical result as
\begin{eqnarray}
<N_K>_{\rm eq}^{\rm C}={[<N_K>_{\rm eq}^{\rm GC}]}^2 \ll
{<N_K>
_{\rm eq}^{\rm GC}}. \end{eqnarray} This shows the
importance of the canonical description of quantum number
conservation when the multiplicity of particles carrying non-zero
$U(1)$ charges is small. We also note that the volume dependence
in the two cases differs. The particle density in the GC limit is
independent of $V$ whereas in the opposite canonical limit the
density scales linearly with $V$. Secondly, we note that the
relaxation time for a canonical system is far shorter than what is
expected from the grand canonical result \cite{ko}. It is also
clear  from Eq.(\ref{5}) that (C) and (GC) limits are essentially
determined by the size of $\langle \delta N^2 \rangle$, the
event-by-event fluctuation of the number of $K^+ K^-$ pairs. The
grand canonical results correspond to small fluctuations, i.e.,
$\langle \delta N^2 \rangle/\langle N \rangle^2 << 1$; while the
canonical description is necessary in the opposite limit.

\section{Model predictions versus experimental data}

In heavy ion collisions the number of produced strange particles
depends  on the collision energy and the centrality of these
collisions. At low collision energies like eg. in GSI/SIS the
freezeout temperature is relatively low being of the order of
$50-80$MeV. Consequently the number of  strange particles in the
final state is very small. Following our discussion in the last
sections it is clear that the statistical description require here
the canonical approach. In Fig. 2 we show the experimental data on
$K^+$ yield per participant $A_{part}$  as a function of
$A_{part}$  measured in Au-Au collisions at $E_{lab} \sim 1$ A/GeV
\cite{data}.
\begin{figure}[htb]
\begin{minipage}[t]{170mm}
{\hskip 4.0cm
\includegraphics[width=21.5pc, height=19.5pc]{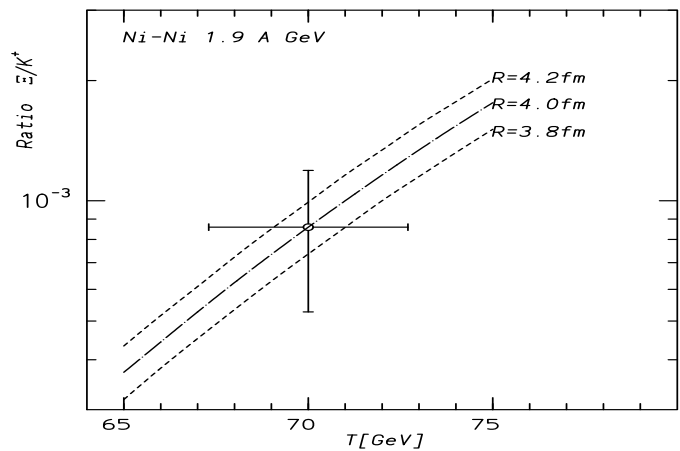}}\\
\vskip -1.0cm {\caption{The ratio of $\Xi$ to $K^+$ multiplicity
as a function of temperature. The point with errors indicates the
predictions of the thermal model for Ni-Ni collisions at 1.9
A/GeV. The lines corresponds to canonical calculations for
different centrality measured by the initial radius $R$ of the
system. }}
\end{minipage}
\end{figure}
The volume parameter in the statistical model scale
with the number of participant. We can thus directly compare the
model with data by fixing  thermal parameters, the temperature and
the baryon chemical potential, such that to reproduce the measured
particle multiplicity ratios \cite{Cle99}. In Fig. 2 the results
of the canonical model is shown by the full line. The results
clearly indicates that strong almost quadratic dependence of kaon
yield on the number of participant is well reproduced by the
model. The quadratic dependence of particle multiplicity in the
canonical regime is the basic  property of the  model. In Fig. 3
we calculate the ratio   $\Xi /K^+$ multiplicities for central
$Ni-Ni$ collisions in the temperature range which corresponds to
the collision energy of $E_{lab}\sim 2$ A/GeV.

 The yield of $\Xi$ is seen to be
substantially smaller then the yield of $K^+$. This result is not
only related with the differences in  particle  masses but
particularly it appears due to the canonical suppression. Since
$\Xi$ carry strangeness minus two it has to be produced together
with two $K^+$ to locally and exactly neutralized  strangeness. It
is also interesting to note that the $\Xi /K^+$ ratio is
independent on the baryon chemical potential. This is because
$K^+$ appears together with $\Lambda$, thus it contains the same
dependence on  $\mu_B$ as multistrange baryons.

 The importance of
the canonical treatment of strangeness conservation is also seen
in higher collision energies like at the SPS when considering
centrality dependance of multistrange baryons. In  peripheral
collisions the yield of strange particles is small such that also
here the canonical description should be applied. The canonical
suppression of thermal particle phase-space increases with
strangeness content of the particle. The exact conservation of
strangeness requires that each particle carrying strangeness $\bar
s$  has to appear e.g. with $s$ other particles of  strangeness
one to satisfy strangeness neutrality condition.

\begin{figure}[htb]
\begin{minipage}[t]{170mm}
{\hskip 4.0cm
\includegraphics[width=21.5pc, height=19.5pc]{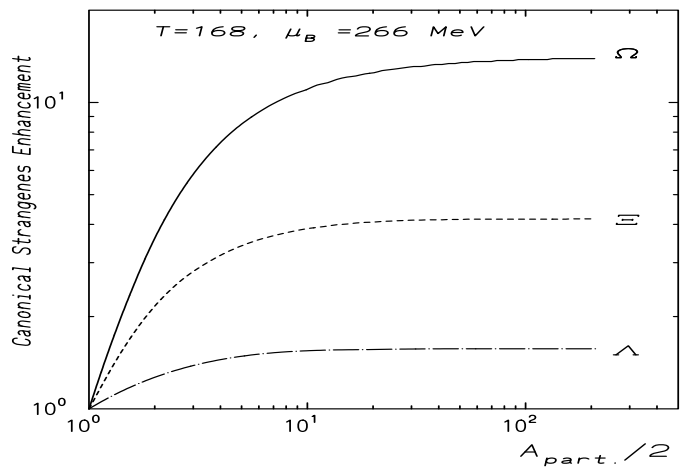}}\\
\vskip -1.0cm {\caption{ Particle multiplicities per participant
normalized to its value in p+p system as a function of the number
of participants  $A_{part.}$  calculated  in the statistical model
in the canonical  ensemble.}}
\end{minipage}
\end{figure}

In Fig. 4 we calculate the multiplicity/participant of $\Omega ,
\Xi ,$ and $\Lambda$   relative to its value in a small system
with only two participants \cite{hamieh}. Thermal parameters  were
assumed here to be $A_{part}$ independent. Fig. 4 shows that the
statistical model in (C) ensemble  reproduces the basic features
of  WA97 data \cite{wa97}: the enhancement pattern and enhancement
saturation for large $A_{part}$ indicating here  that  (GC) limit
is reached. Fig. 4 also demonstrate different $A_{part}$
dependence of strange and multistrange baryons. For small
$A_{part}$ this dependence is power like as describe by Eq.(22).
The quantitative comparison of the model with the experimental
data would require an additional assumption on the variation of
$\mu_B$ with centrality to account for  larger value of $\bar B/B$
ratios in p+A than in Pb+Pb collisions \cite{hamieh,wa97}. The
most recent results of NA57 \cite{na57}, showing an abrupt change
of the enhancement for $\bar\Xi$ are, however,  very unlikely to
be reproducible in terms of the canonical approach.

\section{Summary and conclusions}

We have discussed the importance of the conservation laws in the
application of the  statistical model  to the description of
strangeness production in heavy ion collisions. We have presented
the  arguments that the more general treatment of strangeness
conservation based on the canonical ensemble is required if one
compares the model with experimental data for particle yields
obtained in central A-A collisions at  SIS energies or peripheral
collisions at the SPS. In both situations the number of produced
strange particles per event is still too small  to use the
asymptotic grand canonical ensemble. The time evolution of
strangeness production and the approach to chemical equilibrium
limit was discussed in the context of a kinetic approach. We have
shown on few  examples that the statistical model predictions are
consistent with the experimental data. A more complete
presentation  of the model versus  data can be found in
\cite{stat,Bra99,CLK,Bra95,becattini2,b1,hamieh,Cle99}.

\section{Acknowledgments}
On of us (K.R.) acknowledge stimulating discussions with Su Houng
Lee and the partial support of the  Committee for Scientific
Research (KBN-2P03B 03018). We also acknowledge stimulating
discussions with P. Braun-Munzinger, B. Friman, V. Koch, Z. Lin,
M. Stephanov, H. Satz and X.N. Wang.

\end{document}